\documentclass[aps,pre,longbibliography,notitlepage]{revtex4-1}

\usepackage{bm,amsmath,amssymb,stmaryrd,graphicx,subfigure,afterpage}
\usepackage[abs]{overpic}
\newcommand{\uvc}[1]{\bm{\mathrm{\hat #1}}} 

\newcommand{\bX}{{\bf X}}

\begin{document}

\title{At the end of a moving string}
\author{J. A. Hanna}
\email{hanna@physics.umass.edu}
\author{C. D. Santangelo}
\email{csantang@physics.umass.edu}
\affiliation{Department of Physics, University of Massachusetts, Amherst, MA 01003}

\date{\today}

\begin{abstract}
One cannot pull an open, curved string along itself.  This fact is clearly reflected in the unwrapping motion of a string or chain as it is dragged around an object, and implies strong consequences for slender structures in passive locomotion, whether industrial cables and sheets or the ribbons of rhythmic gymnastics.  We address a basic problem in the dynamics of flexible bodies, namely the solution of the string equations with a free boundary.  This system is the backbone of many fluid-structure interactions, and also a model problem for thin structures where geometric nonlinearities cannot be ignored.  We consider planar dynamics under the restriction that the spatially-dependent stress profile in the string is time-independent, which results in a conservation law form for the equations.  We find a new exact solution whose range of validity is time-dependent, limited to greater than a distance scaling as $t^{\frac{4}{3}}$ from the free end.  The remainder of the distance is covered by splicing to another exact solution for a straight string, which introduces an error into the combined solution.  The splicing also implies a $t^{-\frac{2}{3}}$ singularity in acceleration, apparently corresponding to a whipping motion of the vanishing straight segment.
\end{abstract}

\maketitle

Thin structures interacting with circumambient fluids are associated with beautiful phenomena and applications.
These systems, whether passive or active, locomoting, towed, or tethered, natural or industrial \cite{Gray33, Eloy10, Powers10, ShelleyZhang11}, have long attracted the attention of engineers, physicists, and mathematicians.  An early and important observation was of a towed flexible oil barge buckling and snaking ``like an enraged sea-serpent'' on the Great Ouse \cite{Hawthorne61}. 
The related problem of a pipe conveying fluid is a model nonconservative dynamical system \cite{Nemat-Nasser66, PaidoussisLi93}.  
The rich dynamics and bifurcation landscapes near flat states of pipes, beams, flagella, and flags have been explored over the last several decades.  Yet some thin things spend much time in highly curved states, if strongly forced and weakly resistant to bending.  A full exploration of their physics calls for nonflat solutions from which to make new perturbative excursions, but exact solutions for these problems are scarce.  This remains true even without the analytically daunting complications of nonlocality arising from self-coupling \emph{via} external fluid, or curvature focusing arising from the integrability conditions of two-dimensional surfaces \cite{Witten07}.  There is, therefore, an impetus to uncover exact solutions for the motion of a Kirchhoff rod, Euler elastica, or Routh string \cite{Routh55} under the combined or isolated effects of inertia and local anisotropic drag.  Such solutions are also of general interest, as they will add to the subset of literature on integrable motions of curves that associates a physical origin with the rules of evolution of geometric quantities \cite{Betchov65,Brower84,Cushman-Roisin93,GoldsteinLanger95}. 

We limit ourselves in this paper to strings moving in a plane under the influence of inertia, without external drag.  Far from being a contrived problem, this reduced system of a planar, inertial string is central to the study of rapidly moving tapes, belts, threads, and cables in industrial settings \cite{WickertMote88}, textile manufacture \cite{Mack58,Padfield58,KothariLeaf79both,Fraser92,Clark98}, 
paper processing and printing \cite{Watanabe02}, deposition of oceanic telecommunications cables \cite{Thomson1857-1,*Thomson1857-2,Airy1858,Zajac57}, and the general transport and manipulation of towed payloads in aerial, extraterrestrial, and marine settings \cite{Crist70, SkopChoo71,ChooCasarella72,BeletskyLevin93,SchramReyle68, IversMudie73,MudieIvers75,Sanders82,AblowSchechter83, Dowling88-1}.  Perhaps most importantly, a string is the minimal backbone of any more complicated quasi-one-dimensional thin object where forces may depend on higher order derivatives of position.

The equations of motion for an inextensible, perfectly flexible and twistable string were already clearly formulated over one hundred and fifty years ago.  Despite an early emphasis on curved strings \cite{Routh55}, most solutions known today involve small perturbations around straight strings, including some with free ends and nontrivial stress profiles imposed by gravity, rotation, or external drag \cite{Carrier49, Western80, Belmonte01}.  Some of these are even standard exercises in introductory mechanics texts \cite{FetterWalecka03}.  For nontrivial geometries, a few assaults on the equations of rods and strings have led to approximate or perturbative solutions for slowly varying stress profiles, or near equilibria where spatial stress derivatives are trivial \cite{ColemanDill92, GorielyTabor97-2, McMillenGoriely03, Dickey04}.  Excluding rigid rotations \cite{Mack58,Fusco84,Hanna13}, extant exact, curved solutions are limited to nondispersive waves on strings with spatially uniform stresses.  If the bulk of the string is moving, these solutions fail to satisfy free-end boundary conditions and are thus only valid for infinite or closed-loop strings.  Within this constraint, however, they are valid for arbitrary string shapes and are thus highly generic dynamic equilibria that include steady configurations.

The solutions we currently seek are far from flat and far from equilibrium.  In particular, we wish to treat systems with a single free end, where the bulk of the structure corresponds to a configuration moving primarily, though not exclusively, along its own tangents.  We envision a structure towed along an arbitrary curved path by a vehicle or the wand of a rhythmic gymnast. There exist no known exact solutions for moving curved strings that can accommodate the vanishing of stress at a free boundary.  The equations indicate that this vanishing stress condition implies, through resulting spatial gradients in stress, time-dependence of the geometry.  We will find that, at least for our treatment of the problem, this time-dependence will take the form of a shape propagating towards and reflecting off of the free end.  The motion of a free end during wave reflection can be quite complicated, even in the linear regime \cite{Koh99,Bailey00}.  The related problems of whip cracking \cite{Bernstein58, Krehl98, McMillenGoriely03}, fly casting \cite{Spolek86}, and chains with one fixed and one falling end \cite{Heywood55, CalkinMarch89, Calkin89, Schagerl97, WongYasui06,Tomaszewski06,GeminardVanel08} have been treated approximately by various investigators.  The problem is also likely related to that of deployment from a dense pile \cite{HannaSantangelo12}, where stresses drop rapidly, being strongly screened by curvature.  
 
We also wish to call attention to the phenomenological similarity between the boundary-condition-imposed requirement of end motion in an inertial towed string, and the flutter of pipes, garden hoses, and flags tethered to a static support.  Whether or not this external similarity reflects a deeper physical connection is not immediately clear.  Stresses in tethered structures are thought to arise primarily from drag by an external fluid flow \cite{ShelleyZhang11}, with traveling wave instabilities caused by coupling between this flow and the inertia it generates in the structure.  In a rapidly towed object, significant stresses arise from the inertia of the object itself.  This important distinction only becomes obvious when dealing with generic nonlinear geometries, as the shape-independent stresses generated in the bulk by centripetal accelerations due to primarily tangential motion along a relatively steady curve are not present when one considers a perfectly straight string moving along itself at constant speed.  Thus, our problem is not one that can be treated well by perturbing around a flat shape, as the Galilean invariance of inertia makes this limit singular. It may thus be considered a model problem for geometrically nonlinear approaches.  Indeed, it is well known that the geometrically linearized description of a towed thin object breaks down sufficiently far downstream of the towing point.  For example, Dowling \cite{Dowling88-1} refers to a ``critical point'' on a long towed cylinder where a small-slope description ceases to be valid.  This feature of the geometrically linearized models corresponds to real physical behavior, as evidenced by the use of rope drogues on towed sonar arrays to effectively move the large-slope behavior outside the regions where excessive bending or violent flapping would cause problems with the instrumentation.  Finally, we also note that Moretti \cite{Moretti03} has suggested that inertially-induced tension is important even for tethered objects.

The unsolved problem addressed in this paper, involving the propagation of a shape along a string and its reflection at a free boundary, is both a basic question in the dynamics of flexible bodies and an important behavioral component of a wide class of industrial and natural systems featuring moving thin objects.  We take preliminary steps towards understanding this problem, by finding a new exact solution for the bulk, and constructing an approximate solution for the entire string that satisfies the free end condition.

\section{Notation and Kinematics}

Consider an arc length parametrized, time-dependent, planar curve $\bX(s,t)$ carrying the dyad of unit vectors $(\uvc{t},\uvc{n})$ such that
\begin{equation}
	\partial_s \left(\begin{array}{c}
			\bX\\
			\uvc{t}\\
			\uvc{n}\\
			\end{array}\right) = \left(\begin{array}{cccc}
							0 & 1 & 0\\
							0 & 0 & \kappa\\
							0 & -\kappa & 0\\
							\end{array}\right)
	\left(\begin{array}{c}
	\bX\\
	\uvc{t}\\
	\uvc{n}\\
	\end{array}\right) \, .
\end{equation}
Specification of the curvature $\kappa$, which may take on negative values, defines the curve up to rigid motions.  Locally arc length preserving planar motions may be specified by the tangential velocity $T$,
\begin{equation}
	\partial_t \bX \equiv T\uvc{t} + \tfrac{\partial_s T}{\kappa}\uvc{n} \, ,
\end{equation}
and we may recast the time evolution in terms of the curvature as \cite{Brower84,GoldsteinPetrich91,Nakayama92}
\begin{equation}\label{kappaevol}
	\partial_t \kappa = \partial_s \left(\partial_t\uvc{t}\cdot\uvc{n}\right) = \partial_s\left[ \partial_s \left(\tfrac{\partial_s T}{\kappa}\right) + \kappa T \right] \, .
\end{equation}

\section{Physics}

The physics of a perfectly flexible, inextensible string with uniform mass density $\mu$ is described by the vector wave equation, inextensibility constraint, and boundary condition:
\begin{eqnarray}
	\mu\partial^2_t \bX &=& \partial_s\left(\sigma\partial_s \bX\right) \, , \label{vectorwave} \\
	\partial_s \bX \cdot \partial_s \bX &=& 1\, , \label{vectorconstraint} \\
	\left. \sigma \partial_s\bX \, \right|_{\mathrm{ends}} &=& {\bf{f}}_\mathrm{appl} \, , \label{bc}
\end{eqnarray}
where the stress $\sigma$ is a multiplier field enforcing the constraint \cite{EdwardsGoodyear72,Hinch94,Reeken77,Healey90,Belmonte01,SchagerlBerger02}, and ${\bf{f}}_\mathrm{appl}$ are forces applied to the ends of the chain.  We will be considering a situation where ${\bf{f}}_\mathrm{appl} = 0$ at one end.  In the absence of bending resistance, the curvature and its first derivative are not required to vanish at a free end.

The projections of the arc length derivative of \eqref{vectorwave} along the tangent and normal vectors are
\begin{eqnarray}
	\sigma\kappa^2-\partial^2_s\sigma &=& \mu\partial_t\uvc{t}\cdot\partial_t\uvc{t} \, , \label{tangproj} \\
	2\partial_s\sigma\kappa + \sigma\partial_s\kappa &=& \mu\partial_t^2\uvc{t}\cdot\uvc{n} \, , \label{normproj}
\end{eqnarray}
where we have used \eqref{vectorconstraint} to reduce the order of time derivatives in \eqref{tangproj}, thus also displaying the non-negativity of the expression on the left hand side.  These equations are well known \cite{Routh55} when written in terms of the tangential angle whose arc length derivative is the curvature.  The form of \eqref{tangproj} implies that stresses are screened by curvature and generated by changes in orientation of tangent vectors, whether by centripetal motion of material along the curve, rotation of the curve, or evolution of the curvature.

In two dimensions, $\partial_t\uvc{t}\cdot\partial_t\uvc{t} = \left(\partial_t\uvc{t}\cdot\uvc{n}\right)^2$ and $\partial_t^2\uvc{t}\cdot\uvc{n} =  \partial_t \left( \partial_t\uvc{t}\cdot\uvc{n}\right)$.  Using these relationships along with \eqref{kappaevol}, \eqref{tangproj}, and \eqref{normproj}, we may write
\begin{eqnarray}
	\mu^{\frac{1}{2}}\partial_t\kappa &=& \pm\partial_s\left[\left(\sigma\kappa^2-\partial^2_s\sigma\right)^{\frac{1}{2}}\right] \, , \label{first} \\
	\pm \mu^{\frac{1}{2}} \partial_t\left[ \left( \sigma\kappa^2-\partial^2_s\sigma \right)^{\frac{1}{2}}\right] &=& 2\partial_s\sigma\kappa + \sigma\partial_s\kappa \, . \label{second}
\end{eqnarray}
Equations \eqref{first} and \eqref{second} are, to our knowledge, new.  They are in general quite obnoxious, although for constant stress $\sigma \equiv \sigma_0$ they both reduce to a simple traveling wave expression for the curvature $\kappa$.  Such a constant $\sigma_0$ must be non-negative, according to \eqref{tangproj}.  

On a closed or infinite string, we may ignore the boundary condition \eqref{bc}.  The constant-stress solutions are simply traveling waves:
\begin{eqnarray}
	\partial_t \kappa &=& \pm \left( \frac{\sigma_0}{\mu} \right)^{\frac{1}{2}} \partial_s \kappa \, ,\\
	T &=& \pm \left( \frac{\sigma_0}{\mu} \right)^{\frac{1}{2}} + c_1 \sin \int^s \!\!\!\!d\tilde{s}\,\kappa(\tilde{s}) + c_2 \cos \int^s \!\!\!\!d\tilde{s}\,\kappa(\tilde{s}) \, ,
\end{eqnarray}
with $c_1$ and $c_2$ arbitrary constants that account for Galilean invariance of the system.  Setting these to zero gives us uniform and purely tangential motion with $\sigma_0 = \mu T^2$.  The corresponding shape, designated by $\bX$ or $\kappa$, is motionless in the laboratory frame, and also completely arbitrary, as the wave equation \eqref{vectorwave} is nondispersive for uniform $\sigma$.  These steady solutions of arbitrary shape, reminiscent of lariats \cite{LariatChain,HealeyPapadopoulos90}, were known to Routh \cite{Routh55} through his wrangling with the 1854 Mathematical Tripos, and have also been discovered independently of their mathematical description \cite{Aitken1878,StringLauncher}.

Our discussion has implicitly presumed generic shapes for which $\kappa$ is not simply zero everywhere.  Nonrotating straight lines are special in that their tangent vector fields are also constant vector fields, and thus purely tangential motions of straight lines are also Galilean boosts that leave the stress unaffected.  This makes straight strings singular limits of the towing problem discussed below.

Consider a semi-infinite string, or a string for which one distant end is pulled with the appropriate force to sustain some bulk solution.  Perhaps this bulk motion is lariat-like, with predominantly tangential motion and nearly uniform stress, but the details are unimportant other than that we assume that the tangents are changing and, thus, stresses are nonzero.  We will be concerned with the other, free end $s=0$, where \eqref{bc} tells us that the stress must vanish.  If the stress is continuous, it will have spatial gradients.  These imply, via \eqref{first}, time evolution in shape.  Barring discontinuities, it is impossible to pull an open, curved string so that it moves purely along its tangents.  This may be illustrated quite easily by taking a long string or chain, wrapping it around a smooth convex object, and pulling one end \cite{Cambou12,Calkin89}.  Invariably, the string tries to squeeze or unwrap off of the surface of the object.  Aside from suggesting a nice method for loosening strings tied around frictionless surfaces, this fact implies strong consequences for slender structures in passive locomotion in the absence of guides and obstacles.

We now proceed in search of time-dependent configurations with a free end.

\section{A Solution for Time-Independent Stress}

Hoping for simplification, let's consider stresses $\sigma(s)$ that are independent of time, although still spatially non-uniform.  This transforms the equations \eqref{first} and \eqref{second} into a coupled pair of conservation laws (see also Appendix \ref{cons}):
\begin{eqnarray}
	\mu^{\frac{1}{2}} \partial_t \kappa &=& \pm \partial_s \left( \left[ \sigma\kappa^2-\partial^2_s\sigma\right]^{\frac{1}{2}} \right) \, , \label{conservation1} \\
	\mu^{\frac{1}{2}} \partial_t \left( \sigma \left[ \sigma\kappa^2-\partial^2_s\sigma\right]^{\frac{1}{2}} \right) &=& \pm \partial_s \left( \sigma^2\kappa \right) \, . \label{conservation2}
\end{eqnarray}
These may be combined and integrated to define a function of time $f(t)$,
\begin{equation}\label{relate}
	\frac{\left(\sigma\kappa^2-\partial^2_s\sigma\right)^{\frac{1}{2}}}{\sigma^2\kappa} = \pm e^{f(t)} \, ,
\end{equation}
so that we may invert for $\kappa$,
\begin{equation} \label{kappasq}
	\kappa^2 = \frac{\partial_s^2\sigma}{\sigma\left(1-\sigma^3e^{2f}\right)} \, ,
\end{equation}
and rewrite the conservation laws as
\begin{eqnarray}
	\mu^{\frac{1}{2}} \partial_t \kappa &=& \pm \partial_s \left(e^{f} \sigma^2\kappa \right) \, , \label{cons1} \\
	\mu^{\frac{1}{2}} \partial_t \left( e^{f}\sigma^3\kappa \right) &=& \pm \partial_s \left( \sigma^2\kappa \right) \, . \label{cons2}
\end{eqnarray}
From \eqref{kappasq} and the time invariance of $\sigma$ we may derive
\begin{equation}
	\partial_t \kappa = -\frac{\kappa}{2} \partial_t \ln\left[ \pm \left(\sigma^3e^{2f}-1\right)\right] = \kappa \frac{\sigma^3\partial_tfe^{2f}}{1-\sigma^3e^{2f}} \, ,
\end{equation}
with the sign chosen to make the expression real (for the physically likely situation $\frac{\partial^2_s\sigma}{\sigma} < 0$, the sign should be positive).  As it is derived from what is essentially a compatibility condition for equations \eqref{cons1} and \eqref{cons2}, using this expression to replace $\partial_t \kappa$ in either of those leads to the same thing:
\begin{equation}
	\partial_s\kappa = \kappa \left[ \mu^\frac{1}{2} \frac{\sigma\partial_t\left(e^f\right)}{1-\sigma^3e^{2f}} - \frac{\partial_s\left(\sigma^2\right)}{\sigma^2} \right] \, .
\end{equation}
Thus,
\begin{equation}
	\partial_s\ln\left(\pm\sigma^2\kappa\right) =  \mu^\frac{1}{2} \frac{\sigma\partial_t\left(e^f\right)}{1-\sigma^3e^{2f}} \, ,
\end{equation}
and, using \eqref{kappasq} again,
\begin{equation}
	\pm\partial_s\ln\left(\left[ \frac{\sigma^3\partial_s^2\sigma}{\left(1-\sigma^3e^{2f}\right)} \right]^{\frac{1}{2}} \right) =  \mu^\frac{1}{2} \frac{\sigma\partial_t\left(e^f\right)}{1-\sigma^3e^{2f}} \, .
\end{equation}
Now expand and boil down the left hand side, using the spatial invariance of $f$ in the process, until its denominator matches that of the right hand side, and equate numerators to find:
\begin{equation}
	\pm2\mu^\frac{1}{2}\partial_t\left(e^f\right) + \sigma^2\frac{\partial_s^3\sigma}{\partial_s^2\sigma} e^{2f} - \frac{1}{\sigma}\left( \frac{\partial_s^3\sigma}{\partial_s^2\sigma} + 3\frac{\partial_s\sigma}{\sigma} \right) = 0 \, .
\end{equation}
The only way this equation can hold for non-constant $e^f$ is if the coefficients are all constants.  Let
\begin{eqnarray}
	 \sigma^2\frac{\partial_s^3\sigma}{\partial_s^2\sigma} &\equiv& - B^3 \, , \\
	\frac{1}{\sigma}\left( \frac{\partial_s^3\sigma}{\partial_s^2\sigma} + 3\frac{\partial_s\sigma}{\sigma} \right) &\equiv& C^3 \, .
\end{eqnarray}
Combining gives
\begin{equation}
	\partial_s\sigma = \frac{C^3\sigma^3 + B^3}{3\sigma} \, ,
\end{equation}
which can be solved implicitly.  However, taking derivatives of this equation and reinserting above tells us that $C$ must be zero.  This means that 
\begin{eqnarray}
	\sigma^2 &=& \frac{2B^3}{3}s \, , \\
	e^{2f} &=& \frac{4\mu}{B^6 t^2} \, ,
\end{eqnarray}
for $\sigma(0)=0$, ignoring a constant that would shift time.  Inserting in \eqref{kappasq}, we have
\begin{equation}
	\kappa = \pm \frac{t}{2s} \left[ \frac{1}{\tilde{\mu}s^\frac{3}{2} -t^2} \right]^{\frac{1}{2}} \, ,
\end{equation}
where $\tilde{\mu} \equiv 4\mu\left(\frac{2}{3B}\right)^\frac{3}{2}$.   The curvature $\kappa$ will have the same sign as $\pm e^f$.  Clearly, when $t$ is nonzero there is imaginary curvature at small $s$, which does not correspond to a real rectifiable space curve.  The singularity in curvature is not a problem, as it corresponds to a well-behaved angle.  But we are unable to reach the endpoint $s=0$ except at $t=0$.  To describe the dynamics of the end of the curve will require additional work.

\section{A Splice Job}

We have attempted to reach the end of a generically shaped string, but have fallen a bit short.  However, we have at our disposal other exact solutions that apply for the trivial geometry of a straight string with a free end.  Let us try to splice such an end solution $\bX^e$ to our time-independent-stress bulk solution $\bX^b$:
\begin{equation}
	\bX = \left\{ \begin{array}{rc}
		\bX^e & \quad\quad\, 0 < s < s_m(t) \\
		\bX^b & s_m(t) < s < \infty \quad
	\end{array} \right. \, .
\end{equation}
Influenced by observations of chain dynamics \cite{Cambou12, Tomaszewski06}, we let $\bX^e$ be a straight segment of string that does not rotate but may accelerate along itself.  The length of this end segment will be a function of time, vanishing at $t=0$ when the splice point $s_m(0)=0$ joining the two solutions hits the end of the string.  Unlike that of the bulk piece, the stress in the end segment will be time-dependent.  The stress may, in theory, be kept continuous even at $t=0$, an impossibility for the arrival at the free end of a lariat solution with uniform nonzero stress.  The curvature singularity of the bulk solution does not preclude continuity of tangents of the spliced solution, but such continuity is not a necessity in a perfectly flexible string.

Remarkably, the bulk solution for curvature falls into the rare category of Ces\`{a}ro equations that may be analytically integrated twice to obtain the embedding vector $\bX \equiv X\uvc{x} + Y\uvc{y}$.  Given $\kappa(s,t)$, the tangential angle is $\phi(s,t) = \int^s\!\!ds' \kappa(s',t)$ and the Cartesian coordinates are $X(s,t) = \int^s\!\!ds' \cos \phi(s',t)$ and $Y(s,t) = \int^s\!\!ds' \sin \phi(s',t)$.  The bulk solution $\bX^b$, set horizontal at large $s$, is
\begin{eqnarray}
	\sigma^b &=&  \left(\frac{2B^3s}{3}\right)^\frac{1}{2} \, ,\\
	\kappa^b &=&  \frac{t}{2s} \frac{1}{\left[ \tilde{\mu}s^\frac{3}{2} -t^2 \right]^{\frac{1}{2}}} \, , \\
	\phi^b &=&  - \frac{2}{3}\arctan\frac{t}{\left[\tilde{\mu}s^\frac{3}{2} -t^2 \right]^\frac{1}{2}}  \, , \\
	X^b &=& s\left[ 2\cos\phi^b - \cos\left(2\phi^b \right)\right] \, , \\
	Y^b &=& s\left[ 2\sin\phi^b + \sin\left(2\phi^b \right)\right] \, .
\end{eqnarray}
This will be spliced to the end solution $\bX^e$:
\begin{eqnarray}     
	\sigma^e &=& A(t)s \, , \\
	\kappa^e &=& 0 \, , \\
	\phi^e &=& \phi^e_0(\mathrm{sgn}(t)) \, , \\
	X^e &=& \left[s+ \int^t\!\!\!\!dt' \int^{t'}\!\!\!\!dt'' \frac{A(t'')}{\mu} \right] \cos\phi^e \, , \\
	Y^e &=& \left[s+ \int^t\!\!\!\!dt' \int^{t'}\!\!\!\!dt'' \frac{A(t'')}{\mu} \right] \sin\phi^e \, , \end{eqnarray}
a straight segment that moves along its own tangent.  When this segment disappears at $t=0$, the otherwise constant angle $\phi^e_0$ may change abruptly.  For simplicity of description, we neglect rigid translation and constant terms for the moment.  The forms of $\sigma$ ensure that $\sigma^e(0, t) = \sigma^b(0) = 0$, as the free end condition requires.

The splice point $s_m(t)$ must be greater than the critical value of $s$ that marks the limit of existence of the bulk solution: 
\begin{equation}
	s_m(t)^\frac{3}{2} \ge \frac{t^2}{\tilde{\mu}} \, .
\end{equation}
At this location, a jump condition \cite{Antman05} is imposed:
\begin{equation}\label{jump}
	\left. \left( \sigma^b\partial_s\bX^b - \sigma^e\partial_s\bX^e \right) + \mu \partial_t s_m(t) \left( \partial_t \bX^b - \partial_t \bX^e \right) \right|_{s_m(t)}  = 0 \, .
\end{equation}
Enforcing continuous position leads to the constraints:
\begin{eqnarray}
	&&\begin{split}
	&0=\left[ s_m(t) + \int^t\!\!\!\!dt' \int^{t'}\!\!\!\!dt'' \frac{A(t'')}{\mu}  \right] \cos\phi^e_0 \\
	&+ s_m(t)\left[-2\cos\left(\frac{2}{3}\arctan\frac{t}{\left[\tilde{\mu}s^\frac{3}{2}_m(t) -t^2 \right]^\frac{1}{2}}\right) + \cos\left(\frac{4}{3}\arctan\frac{t}{\left[\tilde{\mu}s^\frac{3}{2}_m(t) -t^2 \right]^\frac{1}{2}}\right) \right]  \, , \label{errorx}
	\end{split}\\
	&&\begin{split}
	&0 = \left[ s_m(t) + \int^t\!\!\!\!dt' \int^{t'}\!\!\!\!dt'' \frac{A(t'')}{\mu}  \right] \sin\phi^e_0 \\
	&+s_m(t)\left[2\sin\left(\frac{2}{3}\arctan\frac{t}{\left[\tilde{\mu}s^\frac{3}{2}_m(t) -t^2 \right]^\frac{1}{2}}\right)  + \sin\left(\frac{4}{3}\arctan\frac{t}{\left[\tilde{\mu}s^\frac{3}{2}_m(t) -t^2 \right]^\frac{1}{2}}\right) \right] \, . \label{errory}
	\end{split}
\end{eqnarray}	
These imply that
\begin{eqnarray}
	s_m(t)^\frac{3}{2} &=& (c^2+1)\frac{t^2}{\tilde{\mu}} \, , \label{sm} \\
	A(t) &=& \frac{4}{9} \mu d^2 \frac{s_m(t)}{t^2} \, , \label{slope}
\end{eqnarray}
for some constants $c^2$ and $d^2$, which relations assure that all the terms have the same $t$ dependence.  As this dependence is not linear in $t$, it cannot be absorbed into a rigid translation term.  This means that the quantities, obtained after insertion of \eqref{sm} and \eqref{slope} in \eqref{errorx} and \eqref{errory}, 
\begin{eqnarray}
	 X^{\mathrm{err}} &\equiv&  s_m(t) \left[ \left(d^2 +1\right) \cos\phi^e_0 - 2 \cos\left(\frac{2}{3}\arctan\frac{1}{c}\right) +  \cos\left(\frac{4}{3}\arctan\frac{1}{c}\right) \right]  \, , \label{bracketx} \\
	 Y^{\mathrm{err}} &\equiv&  s_m(t) \left[ \left(d^2 +1\right) \sin\phi^e_0 + \mathrm{sgn}(t)\left[ 2\sin\left(\frac{2}{3}\arctan\frac{1}{c}\right) +  \sin\left(\frac{4}{3}\arctan\frac{1}{c}\right)\right] \right]  \, , \label{brackety}
\end{eqnarray}	
must both vanish, giving a transcendental relationship between the constants and any possible discontinuity in the tangents.  The jump condition \eqref{jump} provides two more quantities that must vanish,
\begin{eqnarray}
	&&\frac{5}{4}d^2\left(c^2+1\right)\cos\phi^e_0 - \cos\left(\frac{2}{3}\arctan\frac{1}{c}\right) + \frac{c^2+1}{c}\left[\sin\left(\frac{2}{3}\arctan\frac{1}{c}\right) - \sin\left(\frac{4}{3}\arctan\frac{1}{c}\right)\right] \, , \label{jumpx} \\
	&&\frac{5}{4}d^2\left(c^2+1\right)\sin\phi^e_0 + \mathrm{sgn(t)}\left[ \sin\left(\frac{2}{3}\arctan\frac{1}{c}\right) + \frac{c^2+1}{c}\left[\cos\left(\frac{2}{3}\arctan\frac{1}{c}\right) + \cos\left(\frac{4}{3}\arctan\frac{1}{c}\right)\right] \right] \, , \label{jumpy} 
\end{eqnarray}	
obtained after dividing out a common term proportional to $t^\frac{2}{3}$.  This is a total of four equations for three quantities.  We empirically observe, by plotting the functions on the computer, that it appears possible to satisfy all equations by approaching the limit $\phi^e_0 \rightarrow 0$, $c \rightarrow \infty$, $d \rightarrow 0$.  This corresponds to two straight lines with continuous tangents (\mbox{$\phi^e_0 = -  \mathrm{sgn}(t) \frac{2}{3}\arctan\frac{1}{c}$}), one of them having vanishing stress, and an infinite front velocity between the two.  Unfortunately, whether or not it may have a formal justification, this is not a useful limit to take.

Hence, splicing these two exact solutions together in any nontrivial way will lead to an error in the solution, which must be kept small.  For some choice of end angle $\phi^e_0$, there will in general be a finite $c^2$ and $d^2$ that respect the jump conditions \eqref{jumpx} and \eqref{jumpy}.  Using these constants, continuity of position is enforced by subtracting the error terms \eqref{bracketx} and \eqref{brackety} from the end solution $\bX^e$.  The error, localized in the end segment of the resulting ``pseudosolution'', will scale with time in a manner proportional to the true quantities.  A measure of this proportional error is
\begin{equation}\label{properr}
	E \equiv \frac{ \| \mu \partial_t^2 \left(X^\mathrm{err}, Y^\mathrm{err} \right) \| }{ \| \mu \partial_t^2 \left(X^e, Y^e\right) \| } = \frac{1}{d^2}\sqrt{ \left(\frac{X^\mathrm{err}}{s_m(t)}\right)^2 + \left(\frac{Y^\mathrm{err}}{s_m(t)}\right)^2  } \, .
\end{equation}

Neglecting error terms, the magnitude of the acceleration of the end segment is now
\begin{equation}
	\| \mu \partial_t^2 \left(X^e, Y^e\right) \| = A(t) = \tfrac{4}{9}\mu d^2\left(c^2+1\right)^{\frac{2}{3}}  \left(\tilde{\mu} t\right)^{-\frac{2}{3}} \, ,
\end{equation}
while the length of this segment vanishes as $t^\frac{4}{3}$.  The splicing has led to an acceleration of the end piece that goes as $t^{-\frac{2}{3}}$, though the velocity vanishes as $t^\frac{1}{3}$.  The zero-time singularity in acceleration appears to have a physical basis.  The end segment $\bX^e$ swings, with all the other points, around the horizontal straight configuration adopted by the bulk solution $\bX^b$ at $t=0$.  As it vanishes, it experiences a sudden change of angle, as we set $\phi_0^e(-t) = -\phi_0^e(t)$. This change is also the angular phase accumulated by the entire object over the course of the motion.  One could imagine that, in a real system, all of this singular business might be absorbed into some brief, slight stretching of the string.  

Resigned to a pseudosolution, let us try to minimize its error.  It so happens that the relative error $E$ given by \eqref{properr} hovers at a bit more than fifteen percent for jumps in tangential angle of magnitudes less than about $\frac{\pi}{2} -1$, showing little variation in this range but growing large outside it.  The jumps are such as to bend the end piece further away from the bulk piece's straight $t=0$ configuration.  A shallow minimum in error occurs for one such kinked pseudosolution that has a tangent discontinuity of about $\frac{1}{3}$ radian.  

\begin{figure}[here]
\subfigure{
	\begin{overpic}[width=3.2in]{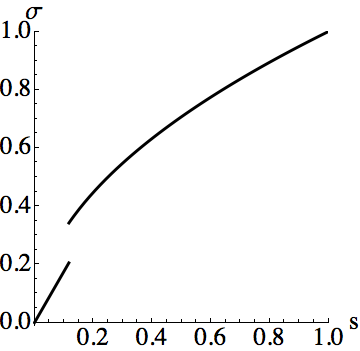}\label{sigmacont}
	\put(150,100){\Large{\subref{sigmacont}}}
	\end{overpic}
	}
\subfigure{
	\begin{overpic}[width=3.2in]{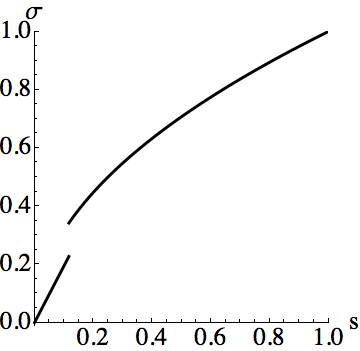}\label{sigmakink}
	\put(150,100){\Large{\subref{sigmakink}}}
	\end{overpic}
	}\\
\subfigure{
	\begin{overpic}[width=3.2in]{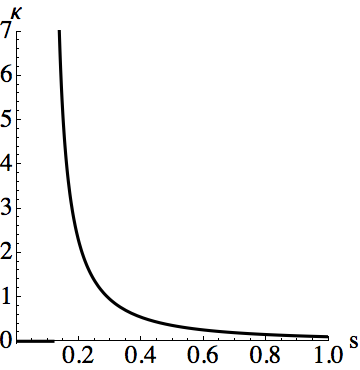}\label{kappacont}
	\put(150,100){\Large{\subref{kappacont}}}
	\end{overpic}
	}
\subfigure{
	\begin{overpic}[width=3.2in]{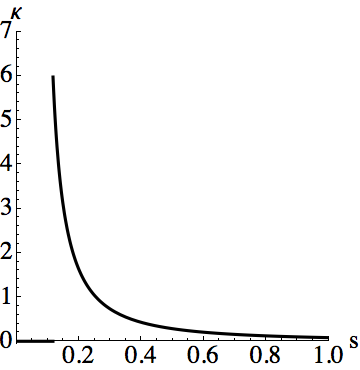}\label{kappakink}
	\put(150,100){\Large{\subref{kappakink}}}
	\end{overpic}
	}
\caption{Snapshot of the stress $\sigma$ and curvature $\kappa$ at $t = \frac{1}{5}\sqrt{\frac{\tilde{\mu}}{c^2+1}}$ with $\tilde{\mu} = \frac{16}{9}$.  \subref{sigmacont} and \subref{kappacont} have parameter values $c=0$, $d^2 = 2.4$ and continuous tangents, while \subref{sigmakink} and \subref{kappakink} have $c = 0.714$, $d^2 = 1.76$, and a jump in tangential angle of $\approx 0.\bar{4}$ radians.  The discontinuities occur at the splice point, where the curvature also diverges for the continuous tangent pseudosolution.  The end segments of the pseudosolutions have a relative error of $\approx \frac{1}{6}$.}
\label{sigmaandkappa}
\end{figure}

Two types of splicing are presented in the figures:  a continuous tangent pseudosolution, and one with a kink of $\frac{4}{9}$ radian, both with $E \approx \frac{1}{6}$.  All examples shown consist of a unit length string, and use $\tilde{\mu} = \frac{16}{9}$ for simplicity.
 Stresses and curvatures at one instant are shown in Figure \ref{sigmaandkappa}.  If tangents are made continuous at the splice point, the curvature there diverges.  The stress is continuous at $t=0$, but not at other times. 

Earlier, a presumed continuity in stress was invoked to explain why strings with free ends must have time-dependent configurations.  Yet our pseudosolution has a stress discontinuity.  One might wonder whether we should have simply attempted to splice a constant-nonzero-stress lariat solution to a zero-stress end solution.  However, for some $\bX^b$ with time-dependent tangents $\partial_s \bX^b \propto \partial_t \bX^b$, one cannot satisfy the jump condition \eqref{jump} with $\sigma^e=0$ and the corresponding constant $\partial_t \bX^e$.  One could, of course, introduce large errors into the end solution, but we have been able to keep the error within a reasonable bound here.  One might also wonder why we don't simply splice two straight strings together to find a solution.  If such a simple approach worked, perhaps it could be extended to multiple strings and, thus, a piecewise approximation to any curve.  In Appendix \ref{straight}, we show that splicing two straight pieces does lead to an exact solution, but a rather unphysical one that also has singularities in velocity and tension at the point of reflection.  Using a different jump condition, an inelastic assumption, and even more severe restrictions on the geometry, Bernstein, Hall, and Trent \cite{Bernstein58} also found discontinuous expressions with velocity and tension singularities.

\begin{figure}[here]
\subfigure{
	\begin{overpic}[width=3.2in]{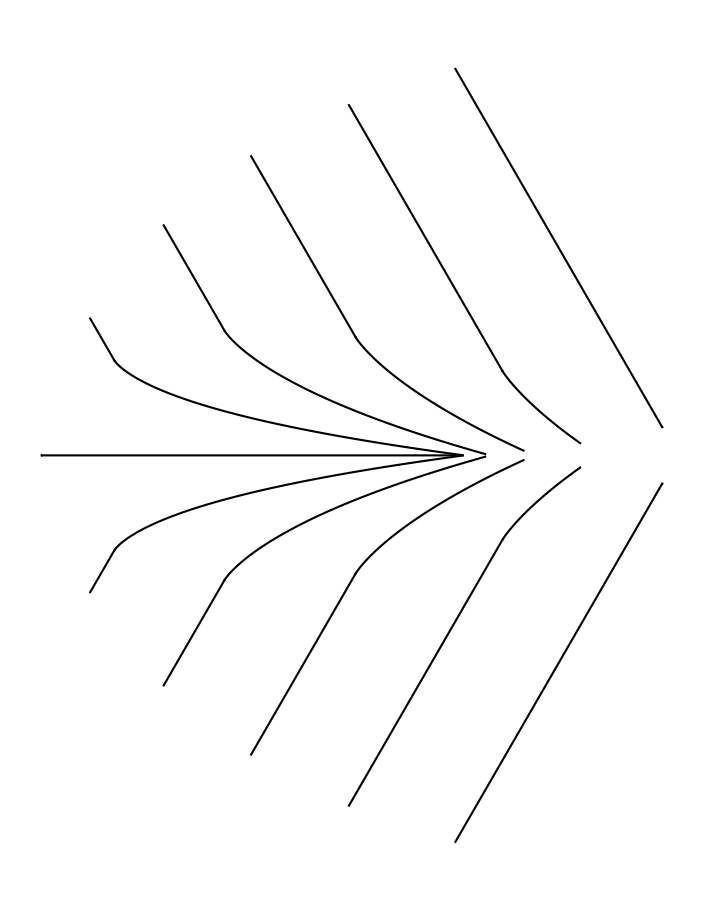}\label{symcont}
	\put(50,250){\Large{\subref{symcont}}}
	\end{overpic}
	}
\subfigure{
	\begin{overpic}[width=3.2in]{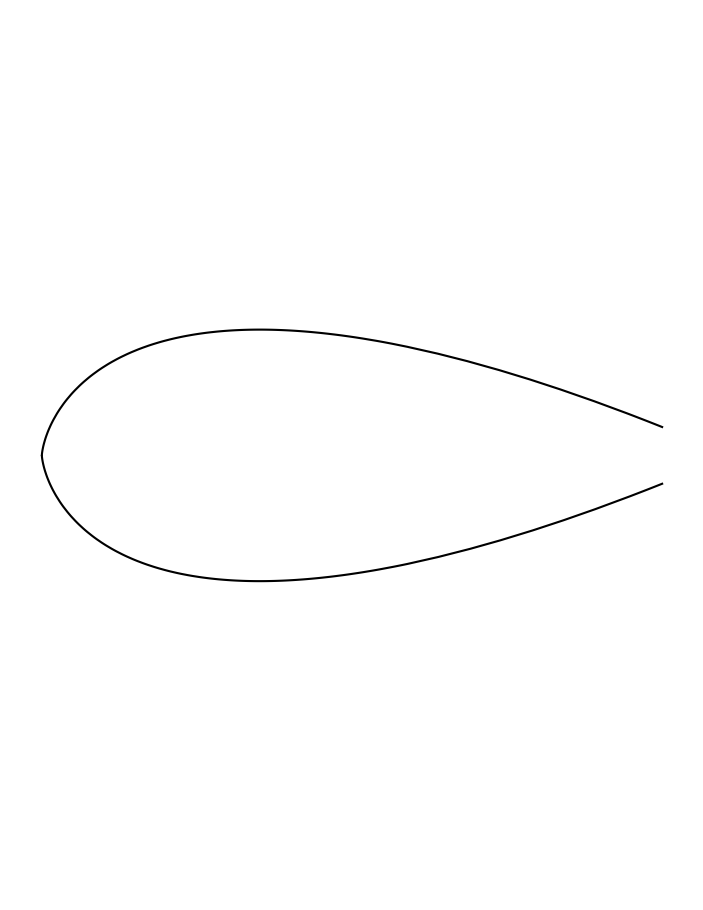}\label{symconttraj}
	\put(50,250){\Large{\subref{symconttraj}}}
	\end{overpic}
	}
\subfigure{
	\begin{overpic}[width=3.2in]{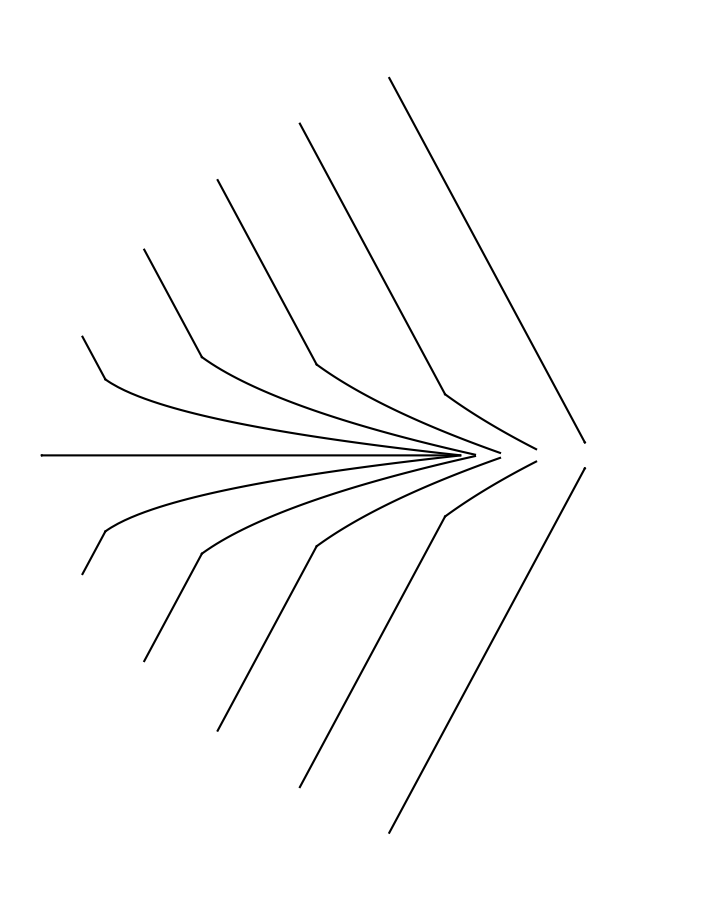}\label{symkink}
	\put(50,250){\Large{\subref{symkink}}}
	\end{overpic}
	}
\subfigure{
	\begin{overpic}[width=3.2in]{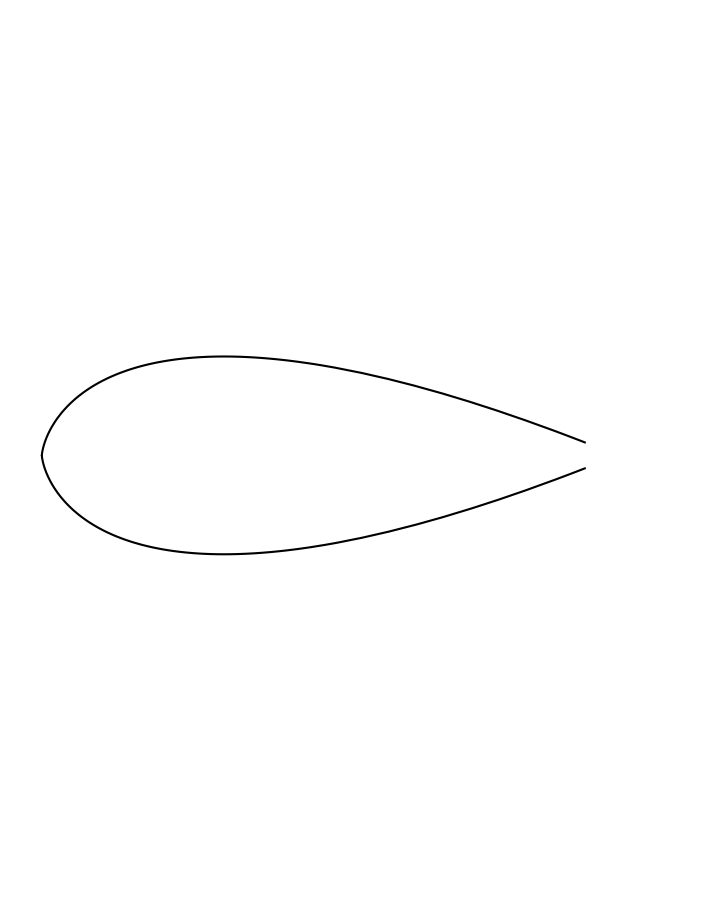}\label{symkinktraj}
	\put(50,250){\Large{\subref{symkinktraj}}}
	\end{overpic}
	}
\caption{\subref{symcont} and \subref{symkink} Snapshots at eleven evenly spaced times between $t = \pm \sqrt{\frac{\tilde{\mu}}{c^2+1}}$ with $\tilde{\mu} = \frac{16}{9}$, and \subref{symconttraj} and \subref{symkinktraj} clockwise trajectories of the splice point, for the pseudosolution \eqref{fisheq}.  The string is of unit length and all images have the same scale.  The upper images \subref{symcont} and \subref{symconttraj} have $c=0$, $d^2 = 2.4$ and continuous tangents; they correspond to the stress and curvature in \ref{sigmacont} and \ref{kappacont}.  The lower images \subref{symkink} and \subref{symkinktraj} have $c = 0.714$, $d^2 = 1.76$, and a jump in tangential angle of $\approx 0.\bar{4}$ radians; they correspond to \ref{sigmakink} and \ref{kappakink}.  The end segments of the pseudosolutions have a relative error of $\approx \frac{1}{6}$.}
\label{sym}
\end{figure}

Uniform velocities may be added to any result.  As $t \rightarrow 0$, $\partial_t Y^b \approx -\frac{8}{3\sqrt{\tilde{\mu}s^\frac{3}{2}}}$, which motion along $\uvc{y}$ may be subtracted out from a favored point for convenience.  Figure \ref{sym} shows snapshots of the configurations, along with the trajectories of the splice points, for the continuous and kinked examples using the pseudosolution
\begin{equation}\label{fisheq}
	\left. \begin{array}{rr}
		\left(X^e - X^{\mathrm{err}} , Y^e - Y^{\mathrm{err}} \right) & 0 < s < s_m \\
		\left(X^b , Y^b   \right) & s_m < s < 1 \;\;\,
	\end{array} \right\} + \left( 0 , \tfrac{8}{3\sqrt{\tilde{\mu}}}  \right) t \, .
\end{equation}
The differences between continuous and kinked are subtle, and the overall angular phase shifts of both curves are $\approx -\frac{2\pi}{3}$.

\begin{figure}[here]
\subfigure{
	\begin{overpic}[width=3.2in]{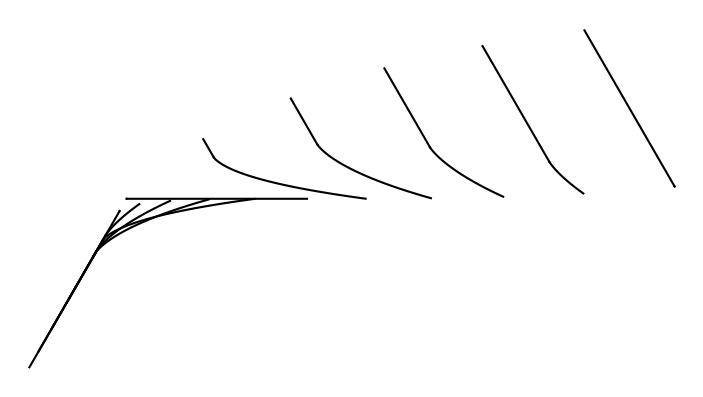}\label{pullcont}
	\put(50,100){\Large{\subref{pullcont}}}
	\end{overpic}
	}
\subfigure{
	\begin{overpic}[width=3.2in]{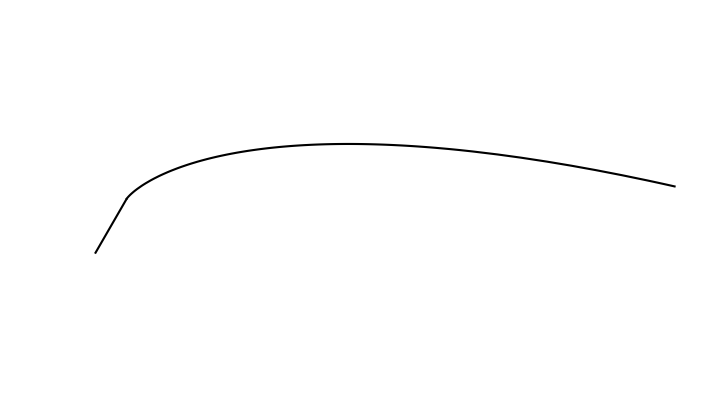}\label{pullconttraj}
	\put(50,100){\Large{\subref{pullconttraj}}}
	\end{overpic}
	}
\caption{\subref{pullcont} Snapshots at eleven evenly spaced times between $t = \pm \sqrt{\frac{\tilde{\mu}}{c^2+1}}$ with $\tilde{\mu} = \frac{16}{9}$, and \subref{pullconttraj} the trajectory of the splice point, for the pseudosolution \eqref{pulleq}.  The parameters are $c=0$, $d^2 = 2.4$, and continuous tangents, corresponding to the stress and curvature in \ref{sigmacont} and \ref{kappacont}.  The string is of unit length and both images have the same scale.  The splice point begins a bit below the kink in the trajectory and initially moves down and to the left, then reverses course for the remainder of the trajectory.
The end segment of the pseudosolution has a relative error of $\approx \frac{1}{6}$.}
\label{pull}
\end{figure}

Adding a velocity along $\uvc{x}$ breaks the symmetry of the patterns.  Figure \ref{pull} shows a continuous tangent example in which the end segment is made to move along itself at early times, by using the pseudosolution
\begin{equation}\label{pulleq}
	\left. \begin{array}{rr}
		\left(X^e - X^{\mathrm{err}} , Y^e - Y^{\mathrm{err}} \right) & 0 < s < s_m \\
		\left(X^b , Y^b   \right) & s_m < s < 1 \;\;\,
	\end{array} \right\} +   \tfrac{8}{3\sqrt{\tilde{\mu}}  \sin  |\phi^e|} \left(\cos |\phi^e| , \sin  |\phi^e|   \right) t \, .
\end{equation}
This example approximately mimics the twin conditions of horizontal pulling at moderate $s$ and tangential motion of the free end near small $s$, a bit like the physical situation of pulling a chain initially wrapped around an obstacle \cite{Cambou12}.

\clearpage

\section{Assessment}

Curved--- that is, generic--- solutions to the string equations with one free and one moving end must have a time-dependent shape.  Our attempt to find such a solution in two dimensions began with the purely pragmatic assumption of time-independent stress, which led to a conservation law form for the equations.  A new solution emerged, exact but with a time-dependent range of validity.  This was extended to the free end by $C^0$ splicing to another exact but geometrically trivial solution, the result being an error in the complete pseudosolution which could be kept on the order of 15\% for moderate jumps in angle at the splice point.  This led to a $t^{\frac{4}{3}}$ scaling for the propagation of information--- the location of the splice point--- to and from the end of the string, and a corresponding $t^{-\frac{2}{3}}$ singularity in acceleration.  The singularity is suggestive of the rapid change in direction of the end segment observed in real systems.

Though it is not a solution, our pseudosolution is less geometrically restricted than prior \cite{Bernstein58} and current (Appendix \ref{straight}) efforts that splice two straight segments together, and avoids the presumably unphysical associated singularities in velocity and tension.  Thus, we may hope that it reflects more information about the real physical system.  Its construction leaves the bulk portion of the solution exact, and satisfies the jump condition at the internal boundary.  However, once the splicing is performed, the formerly exact end portion is no longer a solution, and only approaches a solution in an uninteresting limit.  There likely exists an exact end solution, consisting of some time-dependent curve whose form we do not currently know, that could be spliced to the bulk solution without errors.  There are likely also solutions that do not require splicing of any sort, for which both tangents and stress are continuous, but our results suggest that these will require time-dependent stress in the bulk.

It is not clear how general the present results are.  There is no reason why the stress should be constant in time in the bulk, or why the end of the string should be treated as straight.  The time scalings found here may result from the details of these assumptions.  An important question is whether the effects of the boundary can be localized near the end, or if information really must propagate out from and back into the bulk.  Also open is the question of whether bulk solutions may approach a generic lariat condition of uniform stress, or if the effects of the boundary are truly long-range in this sense as well.  At large $s$, the bulk solution found does not saturate to a uniform value and, even worse, the curvature goes to zero.  It might also be remarked here that, after all our trouble, the curved bulk solutions are not really all that far from straight, and we were further unable to introduce acute angles at the splice point without significant errors.  Future work will consider traveling wave \emph{ans\"{a}tze} for both stress and curvature in an attempt to find a more useful variety of solutions.  

\clearpage

\section*{Acknowledgments}
Work on this problem was inspired by observations made by J.-C. G\'{e}minard and E. Hamm and subsequent discussions with them.  P.-T. Brun and D. Vella shared their unpublished results on a related problem and provided helpful insights.  We also wish to thank S. Arzoumanian and E. de Langre for telling us about relevant literature on towed cylinders.  Funding came from National Science Foundation grant DMR 0846582.

\appendix

\section{Conservation Law Form of the Time-Independent-Stress String Equations in Two Dimensions}\label{cons}

Equations \eqref{conservation1} and \eqref{conservation2} indicate that
\begin{eqnarray}
	\int_{a}^{b} \!\!ds \, \partial_t \kappa &=& \left. \partial_t\uvc{t}\cdot\uvc{n} \, \right|_{a}^{b} \, ,\\	
	\mu \int_{a}^{b} \!\!ds \, \sigma \partial_t\left( \partial_t\uvc{t}\cdot\uvc{n}\right) &=& \left. \sigma^2 \kappa \, \right|_{a}^{b}	\, .
\end{eqnarray}
The first of these is just a kinematical truth following from the definition of curvature as the arc length derivative of the tangential angle.  The second says that the mass density $\mu$ times the total stress-weighted angular acceleration over the length of the string is equal to the difference in a quantity $\sigma^2 \kappa$ at the two ends.  Other conservation laws for rods were discussed in \cite{MaddocksDichmann94}.

\section{Another Splice Job}\label{straight}

Let's consider splicing two straight strings together.  Both bulk and end pieces $\bX^b$ and $\bX^e$ now take the similar forms:
\begin{eqnarray}     
	\sigma^e &=& A^e(t)s \, , \\
	\kappa^e &=& 0 \, , \\
	\phi^e &=& \phi^e_0(\mathrm{sgn}(t)) \, , \\
	X^e &=& \left[s+ \int^t\!\!\!\!dt' \int^{t'}\!\!\!\!dt'' \frac{A^e(t'')}{\mu} \right] \cos\phi^e \, , \\
	Y^e &=& \left[s+ \int^t\!\!\!\!dt' \int^{t'}\!\!\!\!dt'' \frac{A^e(t'')}{\mu} \right] \sin\phi^e \, , 	\\
	\sigma^b &=& A^b(t)s + \sigma^b_0 \, , \\
	\kappa^b &=& 0 \, , \\
	\phi^b &=& \phi^b_0 \, , \\
	X^b &=& \left[s+ \int^t\!\!\!\!dt' \int^{t'}\!\!\!\!dt'' \frac{A^b(t'')}{\mu} \right] \cos\phi^b \, , \\
	Y^b &=& \left[s+ \int^t\!\!\!\!dt' \int^{t'}\!\!\!\!dt'' \frac{A^b(t'')}{\mu} \right] \sin\phi^b \, .
\end{eqnarray}
We can add a linear term to the velocities, but this will not affect our discussion.  By matching positions at the splice point, we find that for generic jumps in tangential angle we must have $A^e(t) = A^b(t)$ and $s_m(t) = - \int^t\!\!dt' \int^{t'}\!\!dt'' \frac{A^e(t'')}{\mu}$.  These relationships are problematic, as they indicate that material points are moving towards the end rather than away from it, and this seems rather unlike the physical picture we hope to describe.
 The jump condition \eqref{jump} now tells us
\begin{equation}
	\sigma^b_0 \left(\begin{array}{c}
				\cos\phi^b\\
				\sin\phi^b\\
				\end{array}\right) +\partial_t\left[ s_m(t) \int^{t}\!\!\!\!dt' A^e(t') \right] \left[ \left(\begin{array}{c}
				\cos\phi^b\\
				\sin\phi^b\\
				\end{array}\right) - \left(\begin{array}{c}
				\cos\phi^e\\
				\sin\phi^e\\
				\end{array}\right) \right] = 0 \, .
\end{equation}
Generic jumps in tangential angle require the coefficients of both vectors to be zero.  Hence, $\sigma^b_0 = 0$ and 
\begin{eqnarray}
	s_m(t) &=& \frac{c^2}{\mu} t^{\frac{1}{2}} \, , \label{sm2} \\
	A^e(t) &=& \frac{c^2}{4}t^{-\frac{3}{2}} \, , \label{slope2}
\end{eqnarray}
with $c^2$ some constant.  This is considerably worse than our previous case, because the velocity blows up as $t^{-\frac{1}{2}}$, the acceleration as $t^{-\frac{3}{2}}$, and now the stress blows up as well.  We are, however, able to specify any angle at the splice point, and the solution is exact.

\bibliographystyle{unsrt}


\end{document}